# How well developed are Altmetrics? Cross-disciplinary analysis of the presence of 'alternative metrics' in scientific publications[1]

Zohreh Zahedi[1], Rodrigo Costas[2] and Paul Wouters[3]

[1a&b] *z.zahedi.2@ cwts.leidenuniv.nl,* [2] *rcostas@cwts.leidenuniv.nl,* [3] *p.f.wouters@cwts.leidenuniv.nl,*

[1a,2, 3] Centre For Science and Technology Studies (CWTS), Leiden University, Leiden, The Netherlands

[1b] Department of Knowledge & Information Sciences (KIS), Faculty of Humanities, Persian Gulf University, Bushehr, 7516913817 (Iran)

**Abstract**
In this paper an analysis of the presence and possibilities of altmetrics for bibliometric and performance analysis is carried out. Using the web based tool Impact Story, we have collected metrics for 20,000 random publications from the Web of Science. We studied the presence and frequency of altmetrics in the set of publications, across fields, document types and also through the years. The main result of the study is that less than 50% of the publications have some kind of altmetrics. The source that provides most metrics is Mendeley, with metrics on readerships for around 37% of all the publications studied. Other sources only provide marginal information. Possibilities and limitations of these indicators are discussed and future research lines are outlined. We also assessed the accuracy of the data retrieved through Impact Story by focusing on the analysis of the accuracy of data from Mendeley; in a follow up study, the accuracy and validity of other data sources not included here will be assessed.

**Conference Topic**
Topic 1 Scientometrics Indicators: criticism and new developments; Topic 2 Old and New Data Sources for Scientometrics Studies: Coverage, Accuracy and Reliability and Topic 7 Webometrics

**Introduction**
Social media are increasingly investigated by information scientists and will remain an important research theme in the near future (Wang, Wang & Xu, 2012). The development and increasing use of the tools has created new challenges for research and many scholars have begun to investigate the impact of social-networking sites on scholarly communication. There is a growing interest in tracking and measuring scholar's activities on the web, through the use, development and combination of new methods and indicators of research with other more traditional impact metrics and web-based alternatives such as webometrics, cybermetrics, and recently social web analysis or Altmetrics (Priem et al., 2010; Wouters & Costas, 2012). Citation analysis is a popular and useful measurement tool in the context of science policy and research management. Citations are usually considered as a proxy for 'scientific impact' (Moed, 2005). However, citation analysis is not free of limitations, and the need for alternative metrics to complement previous indicators has become an object of many studies. Researchers have explored and made use of other metrics (such as log analysis, usage counts, download and view counts, webometrics analysis, etc.) (Haustein, 2012, Thelwall, 2008 & Thelwall, 2012) to overcome the weakness of traditional impact measurement.

An important approach is "altmetrics" which was introduced in 2010 (Priem, et al., 2010) as a novel way of "assessing and tracking scholarly impact on social web", to enhance the process of measuring scholarly performances. In recent years, there has been a growth in the diversity of tools (and also companies) that aim to track 'real-time impact' of scientific outputs by exploring the sharing, reviews, discussions, bookmarking, etc. of scientific publications and sources. Among these tools and companies are F1000 (http://f1000.com/), PLoS article-level-metrics (ALM) (http://article-level-metrics.plos.org/), Altmetric.com (http://altmetric.com/), Plum Analytics (http://www.plumanalytics.com/), Impact Story [2] (http://impactstory.org/), CiteULike (http://www.citeulike.org/), and Mendeley (http://www.mendeley.com/).

---

[1] . This paper presented at the14th International Conference on Scientometrics and Informetrics (ISSI), 16-19 July 2013, University of Vienna, Austria.

**Objectives**

This paper builds upon Wouters & Costas (2012). Our general research objective is to explore whether the new metrics allow for the analysis of more dimensions of impact than is currently possible through citation analysis and what kind of dimensions of scientific activity or performance might be represented by the new web based impact monitors. In exploring these issues, we pursue the following research questions:

1) What is the accuracy and validity of the data retrieved by Impact Story (IS) from Mendeley? Are there any limitations to take into account when using this tool?

2) What is the presence of altmetrics across scientific fields and document types?

3) What is the potential of altmetrics in measuring research performance? What are the relationships between altmetrics and citation indicators?

**Research design and methodology**

We have focused on IS. Although still at an early stage ('beta version'), IS is one of the current web based tools with more potential for research assessment purposes (Wouters & Costas, 2012). IS aggregates "impact data from many sources and displays it in a single report making it quick and easy to view the impact of a wide range of research output" (http://impactstory.org/faq). It takes as input different types of publication identifiers (e.g. DOIs[3], PubMed[4] ids, URLs[5], etc.); which are run through different external services to collect the metrics associated with a given 'artifact' (e.g. a publication); thus a final report is created by IS and shows the impact of the 'artifacts' in different indicators. Using NEW ID () query in SQL, a random sample of 20,000 publications with DOIs (published from 2005 to 2011) from all the disciplines covered by the Web of Science (WoS) has been collected. Using IS, these DOIs were entered into the system and the metrics were collected and saved in CSV format for further analysis[6]. The result table was matched with the CWTS in-house version of the Web of Science on the DOIs (and their altmetric values) to be able to add other bibliometric data to them. Given some mistakes in the table (i.e. missing DOIs from the output coming from IS and also some documents that changed in the meantime in the WoS database) the final list of publications resulted in 19,722 DOIs. Based on this table, we studied the distributions of altmetrics across fields and document types. Citation and collaboration indicators were calculated and the final files were imported in IBM SPSS Statistics 19 for further analysis.

**Analysis of the accuracy of the data retrieved by IS**

In this section we present the result of a manual check on the altmetrics provided by IS, particularly regarding their accuracy with the data from Mendeley. Thus, according to Krejcie,

---

[2] Previously known as Total Impact. For a review of tools for tracking scientific impact see Wouters & Costas (2012), we use IS in this study to refer to Impact Story.
[3] DOI (Digital Object Identifier) is a unique alphanumeric string assigned by the International DOI Foundation to identify content and provide a persistent link to its location on the Internet (http://www.doi.org/)
[4] PubMed comprises more than 22 million citations for biomedical literature from MEDLINE, life science journals, and online books ( http://www.ncbi.nlm.nih.gov/pubmed)
[5] Uniform resource locator (URL) is a specific character string that constitutes a reference to an Internet resource (http://en.wikipedia.org/wiki/Uniform_resource_locator)
[6] Another important element of the data downloaded from IS is that only the first 3 columns (TIID: Total Impact Identifier, Title and DOI) of the CSV files are the same and in the same position, while the other columns are different depending on the values/metrics that they contain (e.g. if in a set of publications only Mendeley and CiteULike metrics are present for the items, only these two columns of metrics would appear, while if in a second search other metrics appear like for example Twitter, then a third column would be added to the field). This situation created the problem that the different files presented a different column distribution, making the merging of all the files in one final table more problematic. The CSV files were uploaded into Google Spread sheets and downloaded back as Plain Text. A SAS program was used to merge all the files together and put them in the template made previously.

& Morgan, (1970), with 95% confidence level, the minimum required sample size for 19722, is 377 observations; therefore, 377 DOIs[7] were selected for manually checking in order to see whether each DOI retrieved refers to the same publication in Mendeley and the same metrics are collected. We found that 208 items had exactly the same scores as before, 154 presented an increase in readerships (which can be explained by the time lag between the download of the data from IS and the manual check) and 4 with decrease in readership counts, 2 items were not found, for 6 items it wasn't possible to get the readership scores and 3 mistaken items were found[8].

Since most of information is entered by users in Mendeley and not all items have DOIs or some may have incorrect DOIs; the title searches of the 377 DOIs were also done in Mendeley in order to see if there are any differences between the DOI and Title search regarding each publication. The result showed that only 10 items can't be found by their titles although they are saved in Mendeley and can be retrieved by their DOIs/Pub Med IDs through IS and only for 2 cases there were metrics through their titles but not through their DOIs. In general, these results suggest that for this sample, the data from Mendeley retrieved through IS is quite reliable although there are some limitations in Mendeley (see Bar-Ilan, 2012)[9] which have to be taken into accounts when checking the data.

**Results and main findings**

Table 1 shows the frequencies and percentage of all altmetrics data retrieved by IS (with the only exception of F1000 that has been left out of this study as they are only available for medical journals and with a yes/no value). Most of the metrics present a very low frequency in our sample, mainly all the PlosAlm indicators as they are only available for the PLoS journals and their presence in our sample is negligible.

**Table 1. Presence of IS altmetrics from all data sources across publications**

| Data Source | papers with metrics | % | papers without metrics | % |
|---|---|---|---|---|
| Mendeley readers | 7235 | **36.7** | 12487 | **63.3** |
| PubMed pmc citations | 2593 | 13.1 | 17129 | 86.9 |
| CiteULike bookmarks | 1638 | **8.3** | 18084 | 91.7 |
| PubMed pmc citations reviews | 929 | 4.7 | 18793 | 95.3 |
| Wikipedia Mentions | 270 | 1.4 | 19452 | 98.6 |
| Facebook likes | 142 | 0.7 | 19580 | 99.3 |
| Topsy Tweets | 95 | 0.5 | 19627 | 99.5 |
| PubMed pmc citations editorials | 55 | 0.3 | 19667 | 99.7 |
| Facebook shares | 57 | 0.3 | 19665 | 99.7 |
| Facebook comments | 42 | 0.2 | 19680 | 99.8 |
| Delicious bookmarks | 33 | 0.2 | 19689 | 99.8 |
| Topsy influential tweets | 18 | 0.1 | 19704 | 99.9 |
| PlosAlm_pmc_full_text | 1 | 0.0 | 19721 | 99.9 |
| PlosAlm _pmc_abstract | 1 | 0.0 | 19721 | 99.9 |
| PlosAlm_pubmed_central | 1 | 0.0 | 19721 | 99.9 |
| PlosAlm _pmc_pdf | 1 | 0.0 | 19721 | 99.9 |
| PlosAlm _pmc_supp_data | 1 | 0.0 | 19721 | 99.9 |
| PlosAlm _pmc_unique_ip | 1 | 0.0 | 19721 | 99.9 |
| PlosAlm _pmc_figure | 1 | 0.0 | 19721 | 99.9 |
| PlosAlm _html_views | 1 | 0.0 | 19721 | 99.9 |
| PlosAlm _pdf_views | 1 | 0.0 | 19721 | 99.9 |
| PlosAlm _scopus | 1 | 0.0 | 19721 | 99.9 |
| PlosAlm _crossref | 1 | 0.0 | 19721 | 99.9 |
| PlosAlm | 1 | 0.0 | 19721 | 99.9 |

---

[7] Altmetrics retrieved by IS contains two parts: metrics found and not found. We decided to select a small sample (two sets of 5 items from each part) to check for the accuracy of data retrieved; therefore both DOIs with and DOIs without metrics were checked
[8] The DOIs retrieved for these 3 were different from the DOIs entered; thus pointing to different articles.
[9] Sometimes, publications can not be found in Mendeley because the titles are not entered correctly by the users; and there are also some duplicates records for a single publication with different number of readerships in Mendeley.

| Data Source | papers with metrics | % | papers without metrics | % |
|---|---|---|---|---|
| Facebook clicks | 16 | 0.1 | 19706 | 99.9 |

Considering table 1, our main finding is that, with the exception of Mendeley, the presence of metrics across publications and fields is very low. Clearly, their potential use for the assessment of the impact of scientific publications is still limited. Based on Table 1, we decided to remove some of the metrics from our study: PlosAlm due to their low frequency and PubMed-based indicators because they are limited only to the Health Sciences and they refer to citations, which we will calculate directly. We also decided to sum the metrics coming from Facebook (i.e. Facebook likes, shares, comments, clicks) given their high correlation (Priem, Piwowar, & Hemminger, 2012) and their relatively low frequency and due to exceeding the downloading limit of IS at the time of data collection, we excluded data from Twitter since it was not reliable.

**The presence of altmetrics across fields**

Figures 1 and 2 show the distributions of altmetrics across major fields of science and document types. The altmetrics presence did not vary much by publication year. Multidisciplinary journals ranked highest in almost all metrics. The major source for altmetrics data in our sample is **Mendeley** with the highest readership from Multidisciplinary fields (55.1% of the publications in this field have at least one Mendeley reader). In **Wikipedia,** Multidisciplinary fields (6.5%) ranked the highest as well.

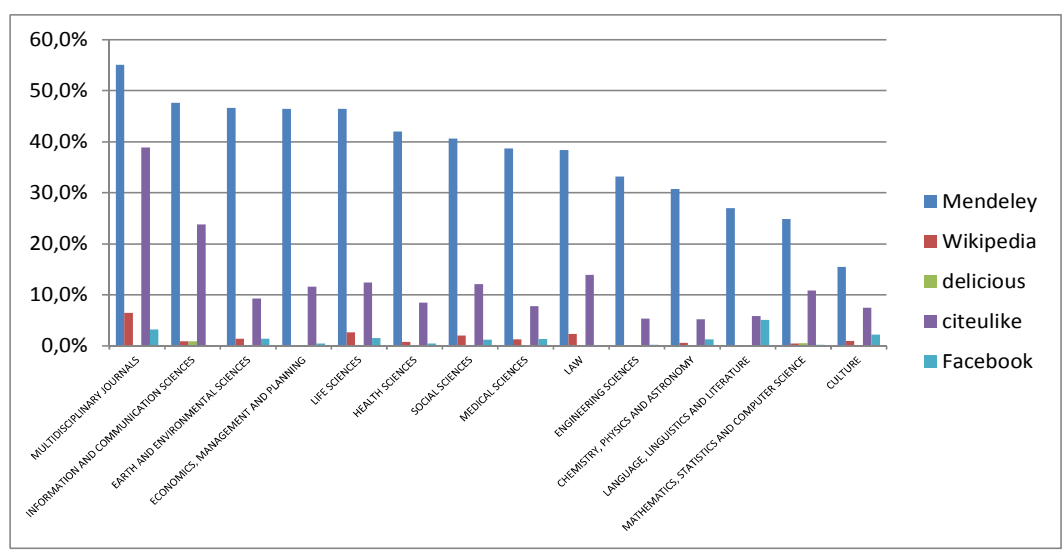

**Figure 1. Distribution of altmetrics across fields**

**The presence of altmetrics across document types**

Regarding document type, there are 16888 (84.6%) articles, 946 (4.79) review papers, 488 (2.47%) letters and 1600 (8.11%) non-citable[10] items in the sample. According to figure 2, around half of (49.6%) the review papers and 40% of articles in the sample have readerships in the Mendeley. With the exception of Delicious, which has a negligible presence, review papers have proportionally attracted more metrics than other document types in our sample, although the number of review papers in our sample is smaller than the number of articles.

---

[10] non-citable document type corresponds with all WOS document types except article, letter and review

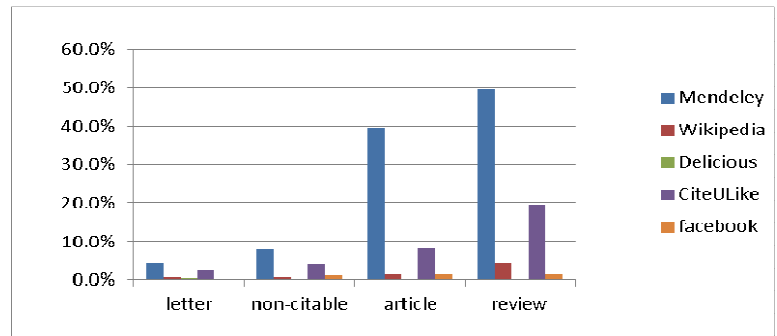

**Figure 2. Distribution of altmetrics across document types**

**Relationships between altmetrics and bibliometric and citation indicators**

In this section we studied the relationship between the main altmetric indicators and citation indicators, publication year, and collaboration indicators (the number of authors and institutions in the papers). For the calculation of the total number of citations we have used a variable citation window (i.e. citations up to the date). Self-citations have been identified for the different publications and introduced in the study as a separate variable. The relationships among altmetric and bibliometric indicators were investigated using Spearman correlation coefficient since the data were skewed (table 2). Concerning citations, we found moderate (r=.3) and small (r=.18) correlations with Mendeley and CiteULike. It is remarkable that Facebook is the source with the lowest correlations with all the other indicators, thus suggesting that this indicator could be related with other types of impact not related to scholarly impact (i.e. measured through citations).

**Table 2. Spearman's Correlation among variables**

|  | *Mendeley readers* | *Facebook* | *delicious bookmarks* | *CiteULike bookmarks* | *pub year* | *Number of references* | *Number of authors* | *Number of institutes* | *Citations* | *self Citations* |
|---|---|---|---|---|---|---|---|---|---|---|
| **Wikipedia mentions** | 0.05 | 0.021 | 0.017 | 0.089 | -0.041 | 0.063 | 0.019 | 0.016 | 0.097 | 0.073 |
| **Mendeley readers** |  | 0.033 | 0.007 | 0.171 | -0.003 | 0.298 | 0.111 | 0.098 | 0.307 | 0.195 |
| **Facebook** |  |  | 0.098 | 0.009 | 0.058 | 0.019 | 0.015 | 0.008 | -0.002 | -0.005 |
| **Delicious bookmarks** |  |  |  | 0.024 | 0.005 | 0.011 | -0.01 | -0.002 | 0.006 | 0.002 |
| **CiteULike bookmarks** |  |  |  |  | -0.015 | 0.152 | 0.003 | 0.033 | 0.185 | 0.119 |
| **pub year** |  |  |  |  |  | 0.045 | 0.034 | 0.034 | -0.431 | -0.268 |
| **n_refs** |  |  |  |  |  |  | 0.142 | 0.149 | 0.407 | 0.313 |
| **n_authors** |  |  |  |  |  |  |  | 0.467 | 0.251 | 0.24 |
| **n_institutes** |  |  |  |  |  |  |  |  | 0.142 | 0.154 |
| **Citations** |  |  |  |  |  |  |  |  |  | 0.692 |

**Conclusions and Discussions**

This study shows that IS, although being in an initial stage of development (it is still in a 'Beta' version), is an interesting source for aggregating altmetrics from different sources. However, we also see important limitations particularly regarding the speed and capacity of data collection and formatting of the data. Out of 19,722 publications 7235 (36.7%) had at least one reader in Mendeley, which is considerably a lower share of Mendeley coverage as compared to previous studies such as 97.2% for JASIST articles published between 2001 and 2011 (Bar-Ilan, 2012); 82% coverage of articles published by researchers in Scientometrics (Bar-Ilan et al., 2012); 94% and 93% of articles published in Nature and Science journals in 2007 (Li, Thelwall and Giustini, 2012); and more than 80% of PLoS ONE publications (Priem et al 2012), followed by 1638 (8.3%) publications bookmarked in CiteULike. Previous studies also showed that Mendeley is the most exhaustive altmetrics data source (Bar-Ilan et al., 2012, Priem et al., 2012). Correlation of Mendeley readerships with citation counts showed moderate correlation (r=.30) between the two variables which is also found in other previous studies (Bar-Ilan, 2012; Priem et al., 2012; and Shuai, Pepe & Bollen, 2012). This indicates that reading and citing are different scientific activities. Multidisciplinary fields (i.e. the field where journals such as *Nature*, *Science* or the *PNAS* are included) attracted more readerships. Review articles were proportionally the most read, shared, liked or bookmarked format compared to articles, non-citable and letters in Mendeley. This may be evidence for the specific role of this document type in dissemination of scientific knowledge. The main result of this study is that the presence of altmetrics is not yet prevalent enough for research evaluation purposes. As indicated in table 1, in our sample, except in Mendeley (63% of publications without metric), in all other data sources more than 90% of the publications are without any metric; thus less than 50% of all publications in this study showed some altmetrics. The amount of altmetrics is still quite low, and given these low numbers problems of validity and reliability could appear when used for real and broad research assessment purposes. For this reason, it is still too soon to consider altmetrics for robust research evaluation purposes, although they already present an interesting informative role. Previous studies also discussed that altmetrics may be useful for the research impact measurement but not proven yet (Li, Thelwall & Giustini, 2012) and in order to be regarded in this context, they need to meet the necessary requirements for data quality and indicator reliability and validity (Wouters & Costas, 2012).

**Acknowledgements**

Special thanks to Erik van Wijk from CWTS for doing the programming part and Jason Priem and Heather Piwowar from Impact Story who gave us very practical and useful hints during the data collection, and special gratitude to Professor Mike Thelwall from Wolverhampton University for his valuable comments on this paper. This work is partially supported by Iranian Ministry of Science, Research, and Technology (MSRT 89100156).

Thelwall, M. (2012). Journal impact evaluation: A webometric perspective, *Scientometrics*, 92(2), 429-441.

Wang, X. W., Wang, Z., Xu, S. M. (2012). Tracing scientist's research trends realtimely. *Scientometrics*, Retrieved 5 January 2013 from: http://arxiv.org/abs/1208.1349

Wouters, P., Costas, R. (2012). *Users, narcissism and control – Tracking the impact of scholarly publications in the 21st century*. Utrecht: SURF foundation. Retrieved September 20, 2012 from: http://www.surffoundation.nl/nl/publicaties/Documents/Users%20narcissism%20and%20control.pdf
7